\documentclass[final,5p,times,twocolumn]{elsarticle}
\usepackage{graphicx,url,amssymb,amsmath,rotating,color,units,wasysym,epsfig,multirow,epstopdf,array}
\usepackage{subfigure}
\usepackage{svg}
\usepackage{float}
\usepackage[colorlinks,urlcolor=blue,citecolor=blue,linkcolor=blue]{hyperref}
\usepackage{soul,xcolor,natbib}
\newcommand{\citeg}[1]{\citep[e.g.][]{#1}}
\setcitestyle{square,aysep={,},yysep={,},citesep={,}}
\hypersetup{pdfauthor={Name}}

\begin{document}
\title{Multi-blade monolithic Euler springs with optimised stress distribution}
\date{\today}
\author{J.V. van Heijningen$^{1, 2,\dagger}$, J. Winterflood$^1$, B. Wu$^{1,3}$, and L. Ju$^{1}$}
\address{$^{1}$ OzGrav-UWA node, University of Western Australia, 35 Stirling Hwy, Crawley WA 6009, Australia}
\address{$^{2}$ Centre for Cosmology, Particle Physics and Phenomenology (CP3), UCLouvain, 1348 Louvain-la-Neuve, Belgium}
\address{$^{3}$ College of Biomedical Engineering \& Instrument Science, Zhejiang University, Hangzhou, China}
\address{\texorpdfstring{$^{\dagger}$}corresponding author: joris.vanheijningen@uclouvain.be}
\vspace{10pt}

\begin{abstract}
Euler springs are used for vertical suspension and vibration isolation as they provide a large static supporting force with a low spring-rate and use minimal spring material. To date, multiple single-width rectangular blades of uniform thickness and stacked flat-face to flat-face have been used in the post buckled state, with half of the blades buckling in each of opposing directions. For ultra-low-noise isolation the ends need to be clamped which results in stick-slip issues at the joints. In this study we investigate the benefits of forming side-by-side oppositely buckling blades from a single monolithic sheet of spring material. Additionally, we study how to distribute the stress evenly along the length of the blade by contouring its width, as well as finding the optimal contour to distribute the stress evenly around the tearing joints between oppositely bending blade sections. We show that this optimal shaping typically improves the inconveniently small spring working range by over 60\% compared to an equivalent rectangular blade.
\vspace{2pc}

\noindent{\it Keywords}: Gravitational waves, Vibration isolation, Euler springs, Seismic noise 
\end{abstract}

\maketitle

\section*{Introduction}
Nowadays, there are many high-technology and research applications where isolation from the ubiquitous seismic motion is required to a high degree. The most challenging of these applications is the endeavour to measure gravitational waves acting on a well-isolated test mass such as done in LIGO\,\cite{aLIGO}, Virgo\,\cite{AdVirgo} and KAGRA\,\cite{KAGRA}.  In this case the motion induced from all noise sources, but particularly seismic, must be reduced from the typical seismic value of order $10^{-10}$\,m/$\surd$Hz around 10\,Hz to below the measurement sensitivity of the instrument which is of order $10^{-20}$\,m/$\surd$Hz.

This level of isolation is achieved by multiple cascaded stages consisting of masses suspended from (blade) springs for vertical isolation, and suspended from pendulum metal wires or glassy fibres for horizontal isolation. The required noise floor dictates that there be no imperfections in the suspension joints which could produce high frequency ($>10$\,Hz) disturbances from low frequency ($<1$\,Hz) servo tracking motions.  Thus there can be no sliding or rolling joints, and all motion must be allowed for by smoothly varying inter-atomic spacing, i.e. flexures and springs.

\begin{figure}[ht]
\centering
\includegraphics[width=0.48\textwidth]{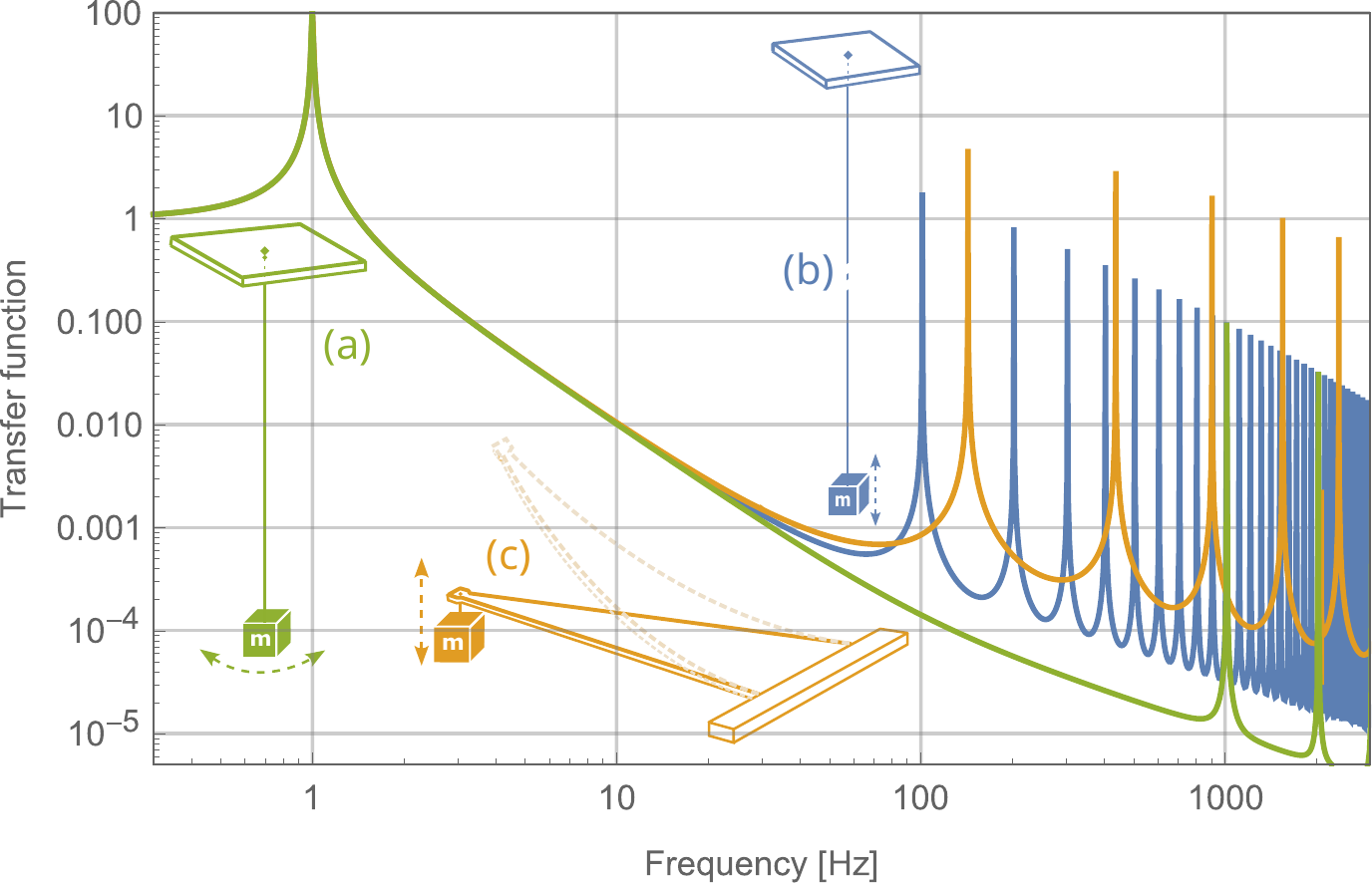}
\caption{(a) Horizontal pendulum isolation compared with (b) stretching fibre and (c) triangular cantilever vertical techniques. All suspension elements made of steel ($E=200$\,GPa), loaded with 10\,kg mass and sized to give 1\% peak strain (2\,GPa stress) and 1\,Hz primary resonance using ANSYS. This required wire $\diameter0.25$, 25\,mm long pendulum, 250\,m long stretching fibre, and a strongly pre-curved (through 210°) cantilever blade 0.8\,mm thick, 147\,mm long by 73.5\,mm wide at the root. Material structural damping was set to $10^{-5}$ and ANSYS' mass damping coefficient adjusted to give a primary resonance Q-factor of 100.}
\label{fig:SuspCompare}
\end{figure}

One of the best isolation techniques for horizontal motion is to suspend the isolated mass with a pendulum fibre. As shown in Fig.~\ref{fig:SuspCompare}, displacement transmissibility falls off with frequency squared above the pendulum natural frequency until internal (violin) modes of the fibre appear. With high-yield-strength pendulum fibres stressed close to breaking, these normal mode frequencies can be three orders of magnitude higher than the pendulum frequency providing a large bandwidth of isolation. In addition, the very small mass of the fibre in relation to the suspended mass means that the violin modes only have a very small effect on the suspended mass.

However, wideband isolation is far more difficult to achieve in the vertical regime. Whereas pendulum isolation in principle requires no energy to be stored in the fibre, material energy storage is a necessity for vertical isolation. If the support moves up and down by a distance $h$ then, for the isolated mass $m$ not to move, $mgh$ of energy must be stored and retrieved from the vertical suspension element. This means that the vertical suspension spring must contain sufficient mass to be able to store this energy as strain.  This unavoidable mass means that the internal modes of the vertical suspension element (e.g. coil spring \emph{surging}) are much lower in frequency and consist of far larger oscillating mass than said pendulum violin modes.  This is demonstrated in Fig.~\ref{fig:SuspCompare}(b) in which the pendulum fibre, already uniformly stressed to some maximum level, is made long enough to achieve 1\,Hz resonance. Now instead of obtaining three orders of magnitude of isolation bandwidth, only two orders can be obtained before isolation is bypassed by a multitude of internal modes.

One of the best vertical isolation techniques is to suspend the mass from the tip of a triangular shaped cantilever blade spring which has been pre-curved so that it becomes straight and horizontal under full load as done in the Virgo superattenuator\,\cite{Beccaria_1997}. The ideal performance of such a spring is also shown in Fig.~\ref{fig:SuspCompare} and it can be seen that its mode-free isolation bandwidth is only slightly (50\%) better than the stretching fibre and the bypass magnitude is many times worse. Thus the energy storage requirement with its related mass requirement should be minimised to obtain the best wideband vertical isolation performance. This remains the prime reason for the development of Euler springs, which are struts or columns with pinned or fixed-ends undergoing elastic buckling.

In this paper, we compare the end-clamped Euler spring approach with a new monolithic spring concept and investigate approaches to improve the blade stress distribution. Section~\ref{EulerIntro} starts with a historical description of non-monolithic Euler spring development and technical grounds for the monolithic blades. In section~\ref{SbsTest} we show test results comparing the clamped end configuration with the same blades applied in tearing joint configuration. In section~\ref{Shape} we determine the optimal main blade shape to achieve an even stress distribution followed by a calculation of its theoretical post-buckle stiffness in section~\ref{ShtoSt}. Section~\ref{PracDes} addresses the issue of high shear stress around the tearing joint where neighbouring strips are bending in opposite directions. We present results of a finite element analysis which attempted to find an optimal filleting radius to prevent the tearing stress from being significantly higher than the main blade bending stress. Section~\ref{Dynamic} compares the simulated dynamic performance of a contoured blade (using finite element modelling) with that of an uncontoured blade of similar dimensions while section~\ref{InstRes} presents semi-static measurements of the new design cut from a glassy metal sample. We end with conclusions and future prospects.

\section{Euler spring development}\label{EulerIntro}

Various styles of spring-mass isolation systems have been developed for the Gingin High Power Optical Facility\,\cite{Dumas_2004,Chin_2006}, Advanced Virgo\,\cite{Beccaria_1997,Cella_2005} and KAGRA\,\cite{Marka_2002}, as well as the LIGO quad suspension \,\cite{Robertson_2002}. Most of these facilities use cantilever springs to obtain vertical vibration isolation. Cantilever springs store a large amount of static vertical suspension energy from their initial loaded deflection in the spring material. This necessitates a large spring mass commensurate with the energy stored. In contrast, the Euler springs~\cite{Winterflood_2002} used in the Gingin isolators store no static energy until loaded with sufficient weight to cause them to buckle. Under buckling they feature low stiffness and relatively high-frequency internal modes due to minimal blade mass. Low stiffness is desirable as it determines the spring-mass resonance and thus the seismic filter cut-off frequency, while high frequency internal modes are desirable as these modes set the frequency above which the seismic filtering becomes ineffective.

An ideal Euler buckling spring force-displacement curve features zero displacement until the force reaches the buckling load, above which it becomes an almost straight line with a gradual slope. Historical work on stability ~\cite{southwell1932analysis} and our previous studies show~\cite{Winterflood_2002c} that the squareness of the knee (buckling point) in the force-displacement characteristic at the onset of buckling is extremely sensitive to the column or blade end boundary conditions. The slightest tendency for the Euler spring blade to buckle out in one direction rather than the other (due to imperfect launching angle, blade curvature, etc. - termed eccentricity) producing a rounded knee with a steep slope and consequent high resonance frequency when operated near the buckling point.

Since sliding or rolling joints are not permitted for this application, the only Euler spring configuration useable is the one with both ends fixed and without any hinged joints.  The  sensitivity to launching angle becomes particularly problematic when these fixed ends of the Euler spring columns or blades are clamped. The reason is that with uniform thickness blades, the regions of highest stress (necessarily a large fraction of yield) occur right where the blades launch from the clamps. The slightest stick-slip motion between the highly stressed blade surface and the differently stressed edge of the clamp produces a non-zero launching angle which spoils the squareness of the buckling knee. Many times we would find that the first buckling cycle had close to ideal square force-displacement characteristic, while subsequent cycles were badly rounded. Since the blade material operated well below yield, we attribute this to stick-slip motion between blade and clamping jaws at the launching point.

The obvious solution to this problem is to use monolithic rather than clamped boundaries at the ends of the blades. The best design would seem to be a step change in blade thickness at the launching point so that clamping would be done on a thick part with low levels of stress and no tendency to stick-slip. Producing a step change in thickness, while possible, presents practical difficulties given that the best spring material comes ready in thin sheets. In this paper we present the design and preliminary results of uniform thickness monolithic Euler spring made from sheet material. The basic approach involves cutting lengthwise slots in a wide sheet to produce multiple side-by-side narrow strips or \emph{blades} which can then be caused to buckle out in opposite directions under lengthwise compression. The strength of the two blade sets which buckle in opposition are matched to give zero average bending moment at the ends. This also makes the squareness of buckling knee insensitive to average launching angle. The highest stress occurs at the \emph{tearing} joints where adjacent blade strips are bending in opposite directions, while the end clamping is some distance away from areas of high stress avoiding stick-slip.

\section{Clamped versus monolithic blades}\label{SbsTest}
\begin{figure}[ht]
\centering
\includegraphics[width=0.48\textwidth]{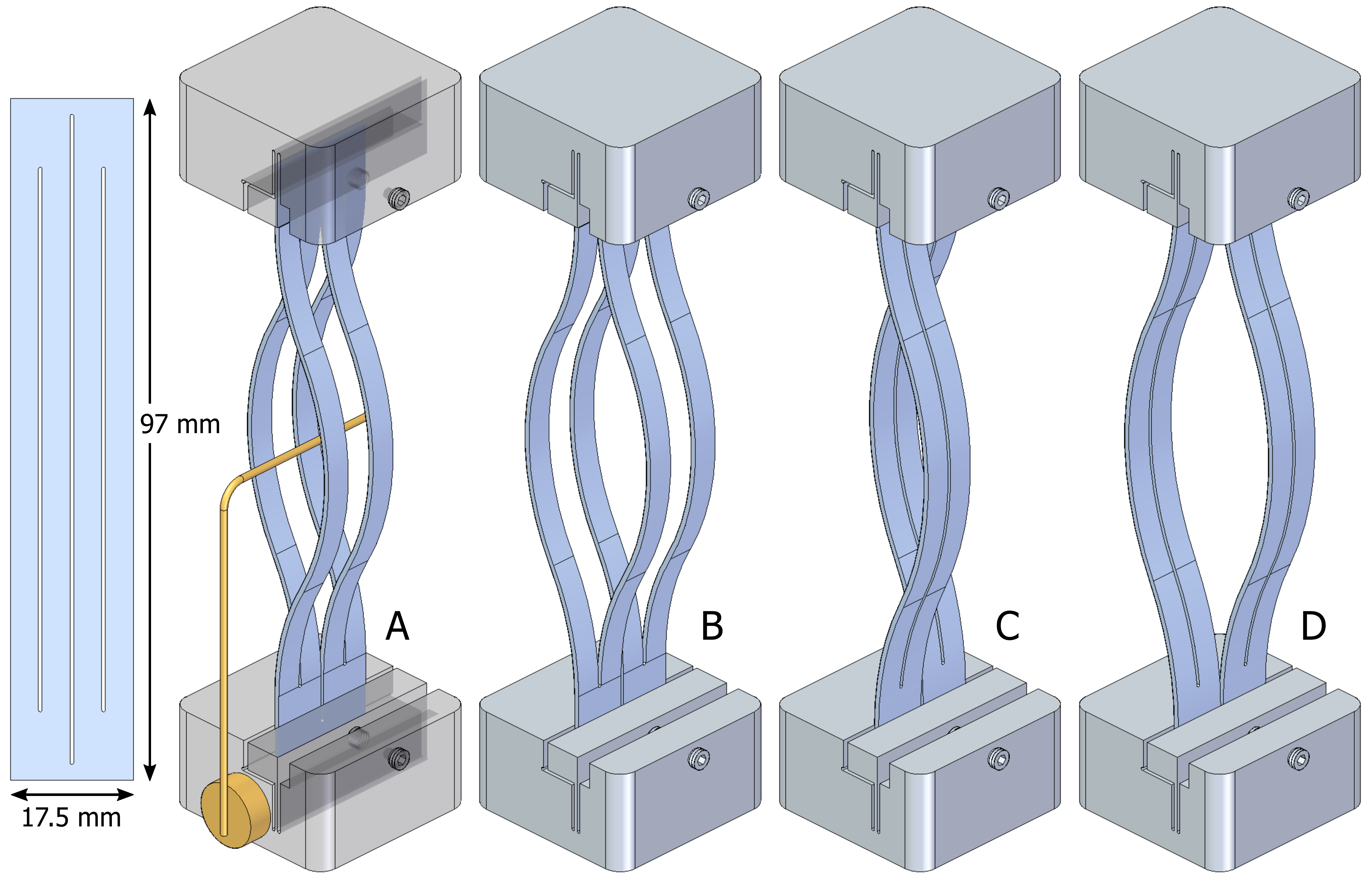}
\caption{Spring blade material was slotted as in the leftmost image and then clamped and compression tested in an Instron 5982 tester with blade bending directions as shown in the other 4 images. Configuration A and B use the monolithic tearing boundary conditions, while configurations C and D rely on clamping. The bent wire in the leftmost image allows the blade buckling directions to be chosen prior to the start of compression cycle.} 
\label{fig:StripConfigs}
\end{figure}

In order to test the effectiveness of the side-by-side monolithic approach, we cut slots of different lengths into a sheet of spring material as shown in Fig.~\ref{fig:StripConfigs}. When the blades are buckled oppositely in pairs (C and D), then the boundary condition at each end of the blades is given by the clamps at the top and bottom. When the blades are buckled in opposite directions singly (A and B), then the boundary condition at each end is given by the tearing joint between the individual blades of each pair and is almost independent of the clamping.

\begin{figure}[H]
\centering
\includegraphics[width=0.48\textwidth]{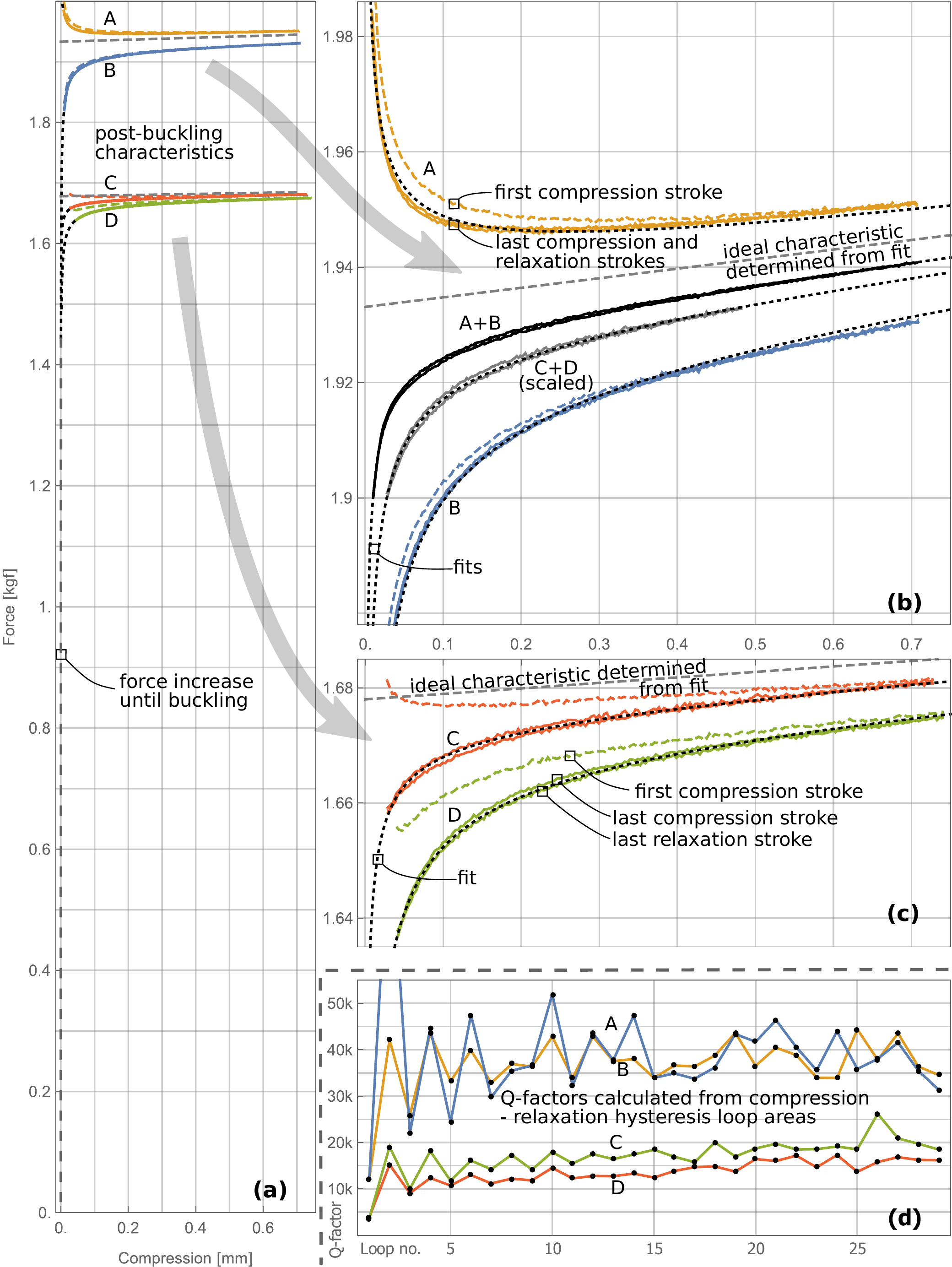}
\caption{Instron compression testing results. (a) Force-displacement characteristics with origin included; (b) Expanded traces for monolithic tearing boundary condition; (c) expanded traces for clamped boundary condition, (d) $Q$-factors calculated from area of each compression-relaxation cycle hysteresis loop.}
\label{fig:WubinResults}
\end{figure}

The spring material used was a carbon steel microtome blade (118$\times$18$\times$0.23\,mm thick from Thomas Scientific, Cat No. 6727C18). This was trimmed and slotted by electric discharge machining to the shape shown on the left in Fig.~\ref{fig:StripConfigs}. The ends were clamped tightly in blocks as shown and mounted in an Instron 5982 tester. Each of the four configurations A through D shown in Fig.~\ref{fig:StripConfigs} were cycled through 15 complete 0.7\;mm compression and relaxation cycles with some of the resulting force-displacement measurements shown in Fig.~\ref{fig:WubinResults}.

Fig.\ref{fig:WubinResults}(a) shows the force-displacement characteristics of the Euler spring tests with the origin included.  It can be seen that they display the typical square knee and almost constant force expected from an Euler buckling spring. The tearing boundary configurations A and B have a slightly shorter bending length and so give higher buckling forces.  The lower traces C and D are the measurements obtained with the standard clamped boundaries.

Mathematical analysis shows that the post-buckling force-displacement relationship can usually be adequately approximated for small displacements with an expression only requiring 3 terms
\begin{equation}
    F = A + B x + C x^{-0.5},
\label{eq:Fit}
\end{equation}
where $A$ is the Euler critical buckling force, $B$ is the post-buckling linear spring coefficient~\cite{Winterflood_2002} and $C$ is an eccentricity factor~\cite{alma9965898902101} with negative values rounding the knee, and positive values sharpening the knee into a cusp. This eccentricity can be due to any non-axisymmetric effect that causes the spring to preferentially buckle in one direction rather than the other.  For clamped flexures this is often due to non-zero average launching angle, but the same effect is observed for initial curvature in the spring or any similar asymmetry such as gravity acting on a horizontal blade.

For each of the pairs A \& B and C \& D a joint mathematical fit using these leading terms was done in order to determine the effective length of the Euler blades (assuming standard Euler post-buckling spring coefficient) and the theoretical buckling force. Although curvature near the knee indicates eccentricity is present, fitting allows this effect to be removed to determine the ideal buckling characteristic. These fits were used to create the dotted black curves appearing through most of the measured data, and the eccentricity term was set to zero to create the dashed gray curves indicating the ideal characteristic had there been no eccentricity. The effective lengths fitted were 58\;mm for A\&B and 88\;mm for C\&D. We do not expect these lengths to correspond very well to the actual blade lengths because the clamped blades are non-uniform (having a slot cut down the middle) and the tearing blade boundaries are far from being simply defined at the ends of the slot.

Fig.\ref{fig:WubinResults}(b)\&(c) are zoomed in views of the post-buckling characteristics for tearing boundary and clamped boundary respectively and are shown at the same scale.  For each case, there are 3 coloured curves--the first compression stroke (dashed line) and the last complete compression-relaxation cycle with colours matching (a), as well as some derived curves in black and gray.

The rounded knee that appears in B is the undesirable characteristic that occurs when the launching angle or inbuilt curvature produces some eccentricity such that that the blade has a preferential buckling direction. This characteristic results in the undesirably higher spring coefficient and resonance frequency that we have often observed with clamped springs after stick-slip has occurred. If the blade is forced to bow in the opposite direction to this structural bias (case A), then a cusp-shaped knee should be obtained instead (which provides a negative spring coefficient or gradient near the cusp).

Since the A and B curves are the monolithic boundary case with no stick-slip launching angle imperfection possible, we conclude that the spring stock must have had some inbuilt differential curvature at the individual blade width scale, which fairly well cancelled itself out at the double blade width scale (as it is not so evident in C and D). The resulting bending direction dependence is apparent in the data from configurations A and B but largely absent for configurations C and D. We expect this is due to residual stresses remaining as a result of the manufacturing, heat treating, or possibly edge sharpening processes. We found this effect was very pronounced in some similar sized microtome blades from another manufacturer that had stamped and indented their logo on one side of their blades!

The first time the blade is compressed to bend in a particular direction (shown dashed in colour), then whatever microscopic grain structures in the blade are poorly constrained, will dislocate easily and semi-permanently for that direction of bend. This produces a relatively large hysteretic loss on the first compression-relaxation cycle, and a non-closed force-displacement characteristic. This effect rapidly reduces for subsequent cycles until steady-state closed hysteresis loops are obtained (shown as solid lines). When the direction of bend is reversed (i.e by switching to configuration B after exercising configuration A), then the same effect appears for the new bending configuration.

With eccentricity present but no hysteretic loss due to stick-slip or yielding, then we expect the eccentricity effect from bending in one direction to be exactly negative of that when bending in the opposite directions (\textit{i.e.} A vs B). Thus if we take the \textit{average} of these two curves (shown in black and labelled A+B), we expect the ideal square buckling knee shown as the dashed gray line. The remaining curvature indicates significant eccentricity in this averaged A+B curve which in this case can only have come from slight yielding around the tearing joints. It is evident from the very small radius of the tearing joints that we had not yet learned how best to limit stress in this area.

For comparison, we have also plotted the average of the C and D curves in gray and labelled C+D. In order for visual comparison, this averaged data has been scaled in amplitude by the ratio of the fitted buckling force, and scaled in displacement by the ratio of the fitted effective lengths (to normalize the compression as a fraction of uncompressed length). It is clear from these plots that the stick-slip losses in the clamped blades (C+D: no yielding expected) are almost twice as bad as the yielding losses in the tearing joint blades (A+B: no stick-slip possible).

It is also evident from the area enclosed by these averaged hysteresis loops in Fig.~\ref{fig:WubinResults}(b), that the energy lost in each compression-relaxation cycle for the clamped (C+D) case is significantly greater than that lost for the tearing (A+B) case. This is quantified in Fig.~\ref{fig:WubinResults}(d). The energy stored and then recovered in each compression-expansion cycle can be calculated quite easily from the measured force-displacement data by integrating the force measured during the compression stroke ($F_c$), and during the expansion stroke ($F_e$). The sum of these values is twice the stored energy while the difference between them gives the energy lost per cycle. Thus the $Q$-factor can be obtained as

\begin{equation}
Q=2\pi\frac{E_{\rm{stored}}}{E_{\rm{loss~per~cycle}}}=\pi\frac{\int F_c\,dx+\int F_e\,dx}{\int F_c\,dx-\int F_e \,dx}.
\label{eq:Qfact}
\end{equation}

This $Q$-factor value can be obtained for any closed hysteresis loop and so may be calculated both for loops which start with a relaxation stroke followed by the compression stroke, as well as the more normal compression-relaxation cycles. Thus 29 $Q$-factor values can be obtained from the 15 compression-expansion cycles as shown in Fig.~\ref{fig:WubinResults}(d). Since the first few strokes do not close properly to form a valid hysteresis loop as the material is still taking on a slight set, the $Q$-factor values obtained for the first few cycles are not strictly valid but are included for interest. This effect appears as the systematic high-low-high-low pattern observed in the early cycles as most of the hysteretic loss occurs in the compression stroke with much less in the return stroke, with the result that the force does not return to its starting value and the loop is not closed. Once this has stabilised, the values obtained become dominated by random scatter.  

Even with mediocre spring material and no consideration given to tearing joint fillet design, the large improvement in the $Q$-factors obtained from the monolithic tearing boundary in comparison with the clamped boundary show clearly that the stick-slip clamping losses that we postulated are almost certainly a major cause of poor Euler spring performance, and reliance on clamping to set the boundary condition should be avoided as far as possible.

\section{Evening out the stress distribution}\label{Shape}
\begin{figure}[ht]
\centering
\includegraphics[width=0.48\textwidth]{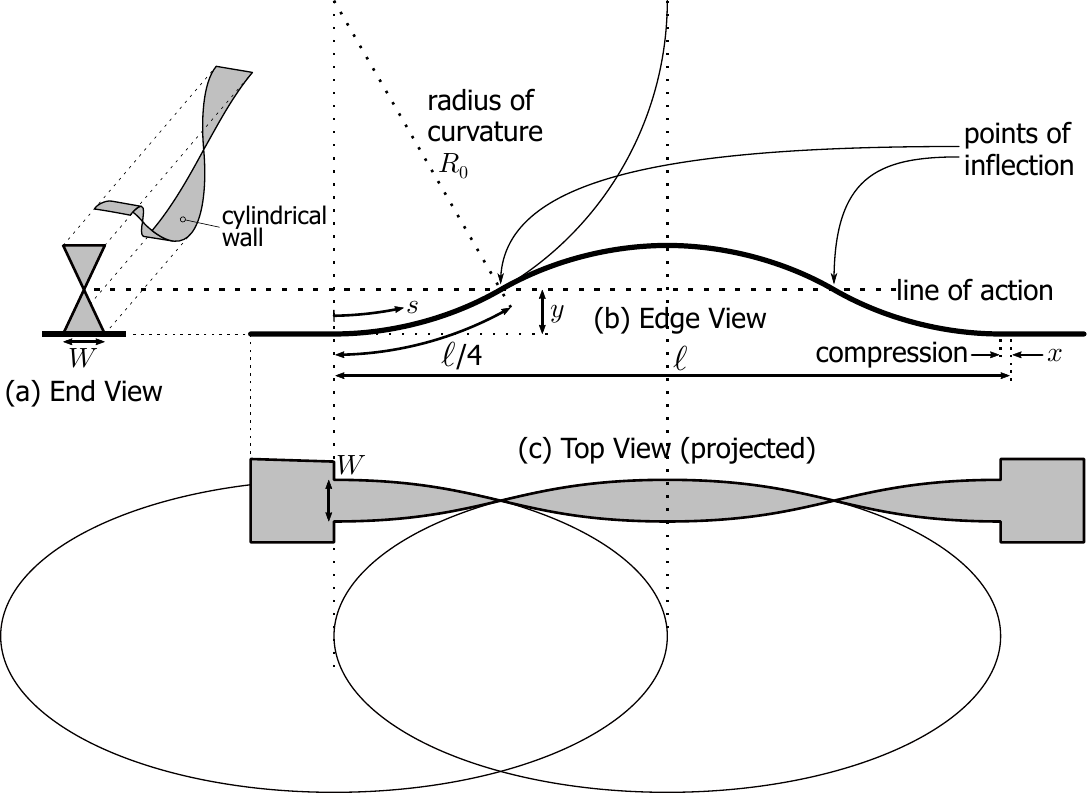}
\caption{A width contoured Euler spring design.  The edge view (b) shows four quadrants of length $\ell/4$ drawn as arcs of circles with radius $R$. The top view (c) in buckled state shows that the resultant shape in projection are ellipses. With no moments appearing at the points of inflection, the blade width becomes zero. The bow-tie end view (a) shows how such a blade would look if one were to observe it along the axis of spring action.}\label{fig:ESprof}
\end{figure}

When a constant width Euler spring with fixed ends buckles, it takes on the shape of an elastica. This bending shape, has quite uneven stress distribution with maximum stress at the ends and midpoint~\cite{Winterflood_2002b}. Consequently other sections of blade that are less stressed remain straighter than they need be resulting in much reduced compression distance (\textit{i.e.} spring working range) than a given stress level should allow.  If the thickness of the blade material is constant, then the material will only become evenly stressed when it is bent to a constant radius---forming segments of circular arcs as shown in the edge view in Fig.~\ref{fig:ESprof}(b).

In order to obtain constant radius of bend, the area moment of inertia should vary in direct proportion to the applied moment. Since points of inflection (PoIs) along the blade mark spots with zero moment, the line-of-action of the applied force must be a straight line through these PoIs and the applied moment must be proportional to the perpendicular distance from this line to the blade. For a constant thickness blade, the moment of inertia is proportional to the width: so the width of the blade should be made directly proportional to its perpendicular displacement from the line-of-action. This proportionality appears as the straight lines of a bow-tie shape in the projected end view in Fig.~\ref{fig:ESprof}(a) where the width of the material is zero at the PoIs and maximum at the points of maximum moment.

Each section of blade that is curved to a constant radius forms part of a cylinder, which when cut by a plane forms part of an ellipse, as indicated in the top view Fig.~\ref{fig:ESprof}(c). When this curved shape is unrolled as flat material, the shape of its edges then form short segments of a sinusoid curve: sections near the sinusoid turning point. Since only small segments near the turning points are used, the blade edge contour is adequately approximated by either the arc of a circle (for CNC manufacturing) or a parabola (for mathematical analysis). The curve changes slightly depending on the degree of maximum buckling as this determines the angle of arc and thus fraction of sinusoidal curve incorporated.  However, with slight-to-medium buckling (\textit{i.e.} $R_0 \geq \ell/2$) and to first order in $R_0/\ell$, the required width $w$ of the blade varies with position $s$ for the first quarter of its length (before the first PoI $0 \leq s  \leq \ell /4$) as a simple parabola:
\begin{equation}\label{eq:prof}
w(s) = W(1-(4s/\ell)^2).
\end{equation}
The remaining three quarters of the blade are simple symmetric reflections of this segment giving the complete shape shown in Fig.~\ref{fig:ESprof}(c) (In projection the exact curve is constructed with segments of ellipses, while unrolled it is constructed with short turning-point segments of sinusoids). Of course the width of any practical blade would not go to zero at the PoIs, but rather have fillets to give some minimum width for robustness, and the curves would be manufactured as circular arcs without detriment.

\section{The impact of blade shape on Euler spring stiffness}\label{ShtoSt}
The stress and strain in Euler spring blade material of thickness $t$ is a simple function of the radius of curvature $R$ to which it is bent. With effective Young's modulus $Y$\footnote{effective modulus for thin material is increased by $1/(1-\nu^2)$, with $\nu$ denoting Poisson's ratio}, the stress $\sigma$ and strain $\epsilon$ peak at $\sigma_0$ and $\epsilon_0$ at the surfaces of the material as
\begin{equation}
    \epsilon_0 = \frac{t}{2R}  ~~~\rm{and}~~~  \sigma_0 = \frac{Yt}{2R}. 
\end{equation}
The actual stress $\sigma$ and strain $\epsilon$ vary from positive maximum at one surface to negative maximum at the opposite surface, linearly through the material. The integral of their product through the thickness gives the energy per unit of bent surface area. With the contoured Euler springs, the radius of curvature $R$ is constant
and the total surface area can be obtained by integrating Eq.~\ref{eq:prof}, we can obtain an expression for the stored energy as \textit{energy per area} $\times$ \textit{area}
\begin{equation}\label{eq:Est}
E_{\rm{st}} = \frac{Yt^3}{24R^2} \times \frac{2}{3}W \ell = \frac{Y I_a \ell}{3 R^2},
\end{equation}
where $I_a$ is the average area moment of inertia from the average width of the blade which is $2/3$ of the maximum width $W$. From the geometry in Fig.~\ref{fig:ESprof}, the relationship between compression distance $x$ and radius of curvature $R$ is
\begin{equation}\label{eq:Comp}
 x = \ell - 4 R \sin{\frac{\ell}{4 R}}.
\end{equation}
Obtaining an inverse series expression for $R^2$ in terms of $x$, substituting this into the stored energy expression (Eq.~\ref{eq:Est}), and differentiating this expression to get the contoured Euler force $F_{\rm{c}}$ as a function of compression $x$ gives
\begin{equation}
    F_{\rm{c}}=\frac{32 Y I_{\rm{a}}}{\ell^2} + \frac{96 Y I_{\rm{a}} x}{5 \ell^3} + \frac{2304 Y I_{\rm{a}} x^2}{175 \ell^4}+\cdots
\label{eq:FxCont}
\end{equation}
The first (constant) term gives the Euler buckling force, the second term (proportional to $x$) gives the initial spring coefficient, while the remaining terms indicate that the force-displacement relationship is slightly non-linear.

The large advantage obtained by ensuring that the entire length of blade is maximally curved in this manner is that the working range of the spring is maximised. For example, if a 60\,mm long contoured blade (without fillets at 1/4 and 3/4 points) is compressed in length by 3.2\,mm, then (from Eq.~\ref{eq:Comp}) its radius of curvature will be 26.3\,mm. A 60\,mm elastica can only be compressed by 1.95\,mm before some parts of the blade are bent to this same peak curvature and stress level. In this case the contoured blades offer a 64\% increase in compression range over a rectangular blade of the same material! Conversely for otherwise similar blades compressed the \textit{same distance}, the contoured version can have significantly lower stress levels ($\sim$22\%) than the rectangular version.

It is of interest to compare Eq.~\ref{eq:FxCont} against the corresponding expression for a standard fixed-fixed rectangular Euler buckling spring. (This may be derived starting from the elliptic integral functions detailed in texts~\citeg{Reddick_1947} that deal with elastica curves) 

\begin{equation}
    F_{\rm{r}}=\frac{4 \pi^2 Y I_{\rm{a}}}{\ell^2} + \frac{2 \pi^2 Y I_{\rm{a}} x}{\ell^3} + \frac{9 \pi^2 Y I_{\rm{a}} x^2}{8 \ell^4}+\cdots
\end{equation}
If we compare contoured against rectangular blades of the same average width (\textit{i.e.} same amount of spring material), then the force needed to buckle the contoured blade is ($8/\pi^2=$) 81\% of that required to buckle a rectangular blade. If we compare contoured against rectangular blades of the same \textit{maximum} width, then the force needed to buckle the contoured blade is ($(2/3)8/\pi^2=$) 54\% of that required to buckle a rectangular blade.

\begin{figure}[ht]
\centering
\includegraphics[width=0.48\textwidth]{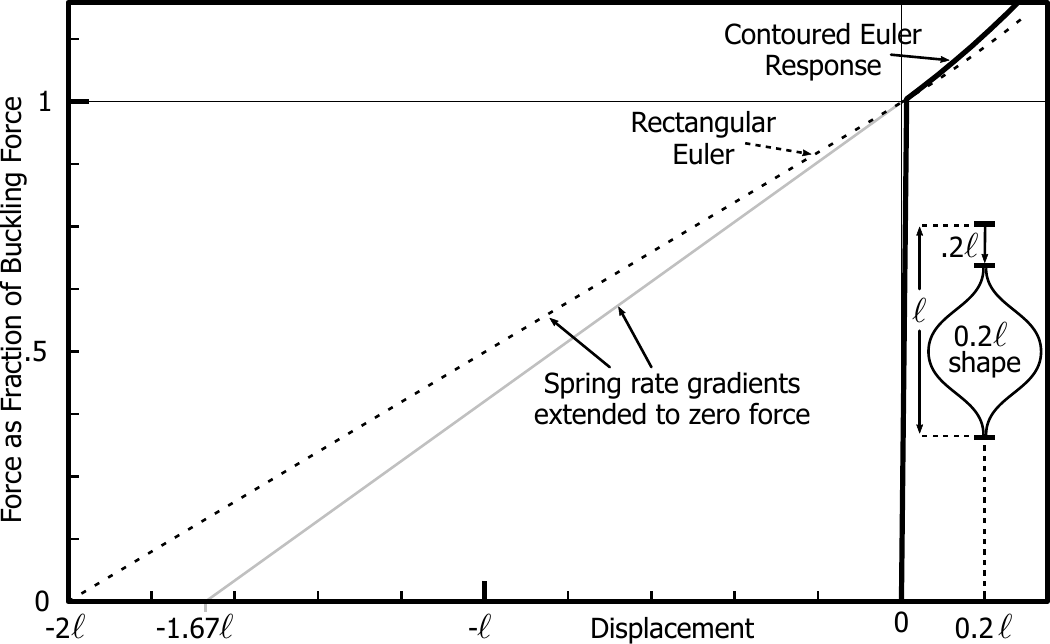}
\caption{The force-displacement characteristic of the contoured Euler spring compared with a rectangular spring. The effective spring length is obtained by extending the initial force-displacement gradient from the buckling point to the zero force axis.}
\label{fig:ForceDisp}
\end{figure}

We usually normalise the spring coefficient to the buckling force, and obtain an equivalent length value (in terms of the unbuckled length of the Euler spring), to which a linear spring with this coefficient would need to be stretched in order to reach the buckling force. This effective length value appears in Fig.~\ref{fig:ForceDisp} as the x-axis intersection when the spring rate gradient is extended backwards from the buckling knee. It can be seen that a linear spring would need to be stretched to twice the length of a rectangular blade to obtain a proportionate spring coefficient. For the case of the contoured blade this value reduces to $5/3$ or 1.67 times the length. Since we would like to obtain large static force with minimal spring coefficient, both the reduction in static force and decrease in effective length are disadvantageous.

Considering the first non-linear spring coefficient term for its fractional change of spring rate with fractional extension indicates that the rectangular spring coefficient increases by 12.5\% per spring length of displacement, while the contoured spring coefficient changes by 37\% per spring length. This non-linearity appears as a curved (rather than straight) line characteristic in the force-displacement plot of Fig.~\ref{fig:ForceDisp}. The contoured spring having 3 times more non-linearity and curvature than the rectangular version. This is slightly disadvantageous as it makes it more difficult to reduce the spring coefficient with negative spring cancellation as the degree of cancellation varies more with displacement.

\begin{figure}[ht]
\centering
\includegraphics[width=0.48\textwidth]{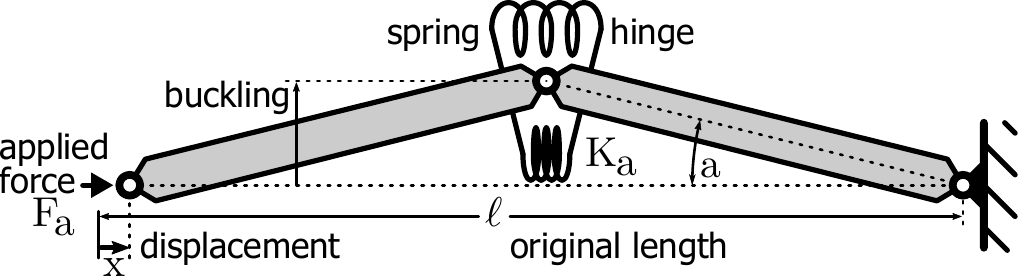}
\caption{A buckling spring in which all the spring energy is stored as torsion in the hinge}
\label{fig:KnuckleBuckle}
\end{figure}

It is interesting to investigate why these parameters have moved in a disadvantageous direction as a result of contouring, and what buckling shape might provide an advantageous change instead. It may be noticed that the main difference between the rectangular and contoured Euler designs is that the strain energy has been spread further away from places of maximum bending (that might be regarded as bending joints) so that with the circular arcs there are scarcely \textit{joints} as such because the bend is evenly distributed all the way along the blade. So one might try the opposite extreme by providing hinged joints so that the bending is maximally concentrated into single spots instead of being distributed along the blade. This puts all the strain energy into torsion springs at the hinged joints. A conceptual arrangement which achieves this is shown in Fig.~\ref{fig:KnuckleBuckle}.

A simple analysis of the force displacement characteristic of this articulated structure gives the following result for low order terms in $x$:
\begin{equation}
    F_{\rm{a}}=\frac{4 K_{\rm{a}}}{\ell} + \frac{4 K_{\rm{a}} x}{3 \ell^2} + \frac{8 K_{\rm{a}} x^2}{15 \ell^3}+\cdots
\label{eq:Fa}
\end{equation}
Repeating this analysis with some small eccentricity (in the form of an initial offset angle) shows that the effect of eccentricity is an additional power series in $x/\ell$ with powers $-0.5, +0.5, +1.5 \cdots$.  Thus for small displacements the force-displacement characteristic of all these types of buckling springs, including ellipticity effects, can be adequately approximated using Eq.~\ref{eq:Fit}.

If we compare these three buckling structures normalized by their critical buckling load and unbuckled length $\ell$ we obtain effective lengths and leading non-linear components as shown in table~\ref{table:EffLen}.

\begin{table}[h!]
\begin{center}

\label{table:EffLen}

\begin{tabular}{ l c c }
Buckling Shape & Eff. length & Non-lin. \\
 \hline \hline
Circular arc bending spring & $1.67\ell$ & $37\%/\ell$ \\

Elastica bending spring & $2\ell$ & $12.5\%/\ell$ \\

Articulated with pivot springs &  $3\ell$   & $40\%/\ell$  \\
\hline
\end{tabular}
\caption{Comparison in effective length and non-linearity for several buckling spring varieties.}
\end{center}
\end{table}

Concentrating all the energy storage towards the conceptual bending joints, does indeed provide a significantly more constant force-displacement characteristic (effective length of 3 instead of 2 or 1.7), but the magnitude of the leading non-linear term in this articulated structure is higher than either of the flexible structures.

\section{Practical blade design}\label{PracDes}
To obtain lateral stability in an Euler spring under compression, the blade width at the PoIs needs to be several times the blade thickness (to ensure it buckles out of the blade plane rather than Z-bending in-plane). This requirement influences the choice of number of blades in that fewer wider blades are more stable against in-plane bending than many narrow blades would be. We chose a symmetric three-bladed design with the central blade having twice the width of the two outer blades to leave no net bending force required of the clamps. Our 3-blade test shapes had hour-glass neck width to thickness ratios of 1:0.75\,mm for the side legs and 2:0.75\,mm for the central blade. In addition the 3-blade arrangement allows the buckling direction to be easily set with a short rod bridging between blades as appears in Fig.~\ref{fig:Photos}(b).

\begin{figure}[ht]
\centering
\includegraphics[width=0.48\textwidth]{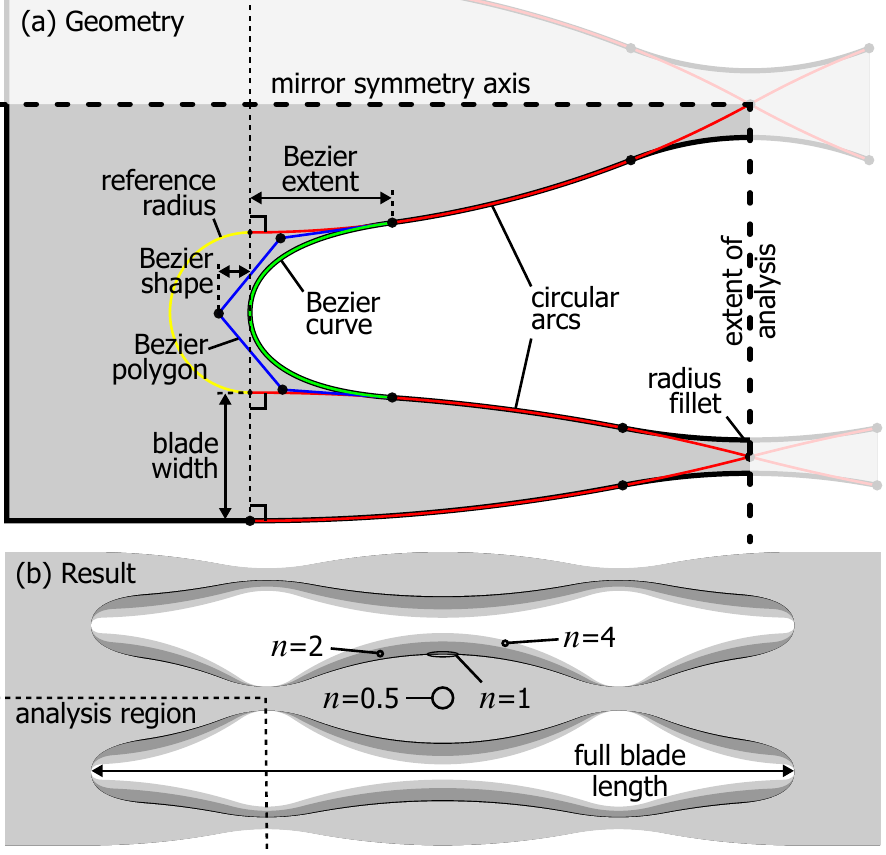}
\caption{Geometry used and result obtained from Ansys shape optimisation. (b) shows the complete 3-blade spring shape with the small analysis region expanded in (a). Three parameters are optimised: blade width, Bezier extent, and Bezier shape; with the optimal result shown in (b). See discussion of Fig.~\ref{fig:OptValPlots} for $n$ parameter}
\label{fig:AnsysGeom}
\end{figure}

It is important to also address the stress levels around the tearing joints between adjacent blades. It is clear that if the joint between neighbouring blade ends is a sharp \textbf{$\vee$} shape, then stresses at the root of the  \textbf{$\vee$} will exceed the main bending stresses as evidenced by \textit{tearing} before bending in weak rigid materials. If instead the tearing joint is a very wide  \textbf{$\cup$} shape compared with the blade width, then the narrow blades must yield in bending before there is any tendency to yield in shear along the wide tear joint. However, this modification reduces the buckling force produced as there is less material (having been removed to give wide  \textbf{$\cup$} joints) resulting in a reduced loading capability. We therefore conducted an extensive optimisation using ANSYS finite element modelling to discover the tearing region shape that gives as much force as possible for a given level of evenly distributed peak stress around the joints. The model analysed is shown in Fig.~\ref{fig:AnsysGeom}.

In order to make the stress distribution as homogeneous as possible, we chose three shape parameters (blade width, Bezier extent and Bezier shape as shown in Fig.~\ref{fig:AnsysGeom}(a)) that could be automatically adjusted by ANSYS to maximize the buckling force while minimizing the peak stress in the structure.  Assuming an array of similar side-by-side blades, the main parameter determining the force required to buckle the structure is the fraction of the rectangular material width that is contoured to form the hour-glass bulge at the midsection of each blade. The wider the gaps between blades (through removing material), the lower the buckling force and operating load but the more room there is to distribute the stress around the tearing joints. Given the blade width at its widest points, the rest of the blade shape is largely dictated as in section~\ref{Shape}, by simple arcs which intersect at the PoIs and meet the end boundary of the blades at right angles as shown by the red lines in Fig.~\ref{fig:AnsysGeom}(a).

We assumed that an optimally even stress distribution would be obtained with a smooth curve filleting the tearing joint, and chose a Bezier curve as shown by the green line in Fig.~\ref{fig:AnsysGeom}(a) with the Bezier control polygon as the blue line. With the Bezier constrained to have its turning point tangential to a suitable point on the blade end boundary-line (dashed), and to join the blade arc (red lines) smoothly, its shape becomes fully constrained  with only two parameters. The first parameter (''extent`` in Fig.~\ref{fig:AnsysGeom}(a)) gives the fraction along the blade length where the (red) arc curve of the blade diverges into the Bezier. The second parameter (''shape`` in Fig.~\ref{fig:AnsysGeom}(a)) is the distance of the mid-Bezier control point (blue) from the blade end boundary line (dashed). This shape parameter determines how rectangular  \textbf{$\sqcup$} through rounded  \textbf{$\cup$} to sharp  \textbf{$\vee$} the filleting is.

Since the entire spring has two axes of symmetry, at most only one quarter of its area needs to be analysed.  With the added consideration of the approximate point-symmetry across PoIs only one eighth of the entire spring needed to be analysed, as shown in Fig.~\ref{fig:AnsysGeom}(a) by the area within the dashed lines. The left hand edge was fixed, and loads applied in steps to the PoI edges to obtain 0.8\,mm of buckled compression without any initial bending imperfection or buckling direction bias. This configuration represents an overall compressing of the 3-bladed spring by approximately 3.2\,mm with adjacent blades buckling out of plane in opposite directions.

\begin{figure}[ht]
\centering
\includegraphics[width=0.48\textwidth]{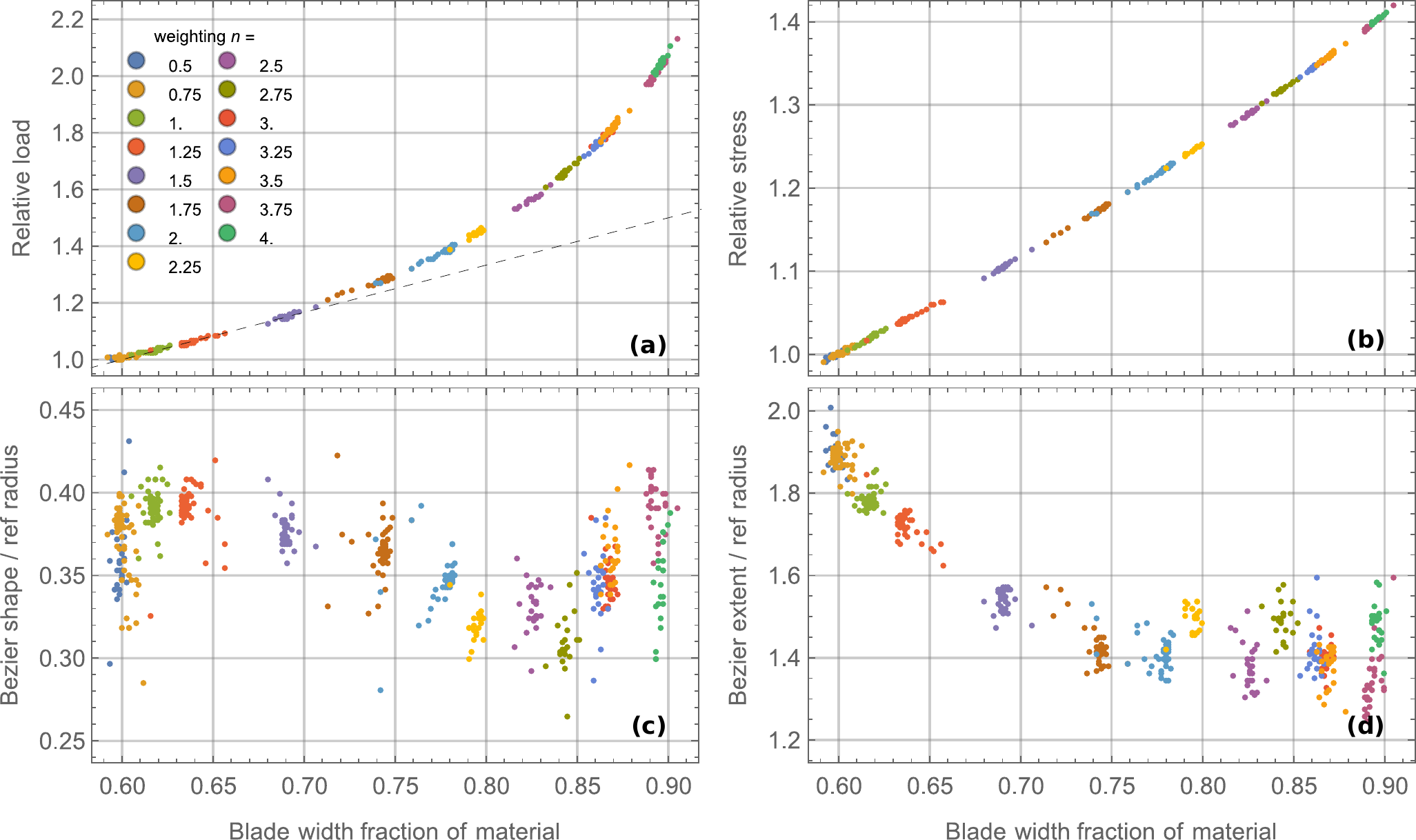}
\caption{Results from FEA optimisation. All results are plotted as a function of the fraction of material width remaining after cutting the slots between blades. Direct proportionality of buckling force (load) with remaining material width is shown as the dashed line in (a) to indicate the significant load advantage obtained with extra width. The stress de-rating factor required for non-optimal tearing joint geometry is shown in (b), while (c) and (d) provide the optimum Bezier ``shape'' and ``extent'' parameter values to choose for any blade width fraction.}
\label{fig:OptValPlots}
\end{figure}

As we want to maximise post-buckle force while minimising peaking of the stress in any small area, we chose $F_{\rm{load}} / \sigma_{\rm{max}}$ as our figure of merit or ANSYS ``objective function''. Since the force obtained from a given width of material may be more important for some applications than the stress level reached, we introduced the parameter $n$ in our objective function as $F_{\rm{load}}^n / \sigma_{\rm{max}}$.  This parameter determines the weighting placed on maximising force at the expense of less than minimum peaking of stress. We varied $n$ in steps of $0.25$ from $n=0.5$ to $n=4$ and obtained 78 ``design points'' (ANSYS terminology) for each value of $n$. The full results of these optimisation runs are plotted in Fig.~\ref{fig:OptValPlots} while the typical shapes obtained are displayed in Fig.~\ref{fig:AnsysGeom}(b).

With $n$ less than unity (making stress evenness more important than force) we found that almost identical shapes were obtained irrespective of $n$. This can be seen in Fig.~\ref{fig:AnsysGeom}(b) where the only difference between the optimum shape for $n=0.5$ (the innermost grey shape) and $n=1$ is the very fine black line bordering the $n=0.5$ inner grey shape. This optimal shape kept approximately 60\% of the material width as blades and $\sim$40\% was cut away to form the gentle fillets in the tearing joints. When $n$ was raised above unity, then the width of the blades started to increase (up to 90\% for $n=4$), and the peak stress started to gather evenly around the tearing joint rather than appearing with the same intensity along the blades. Thus it is clear that the shape obtained for $n\leq1$ is the optimum for even distribution of stress between the blades and the tearing joints, and there seems no point in having gaps between blades larger than this.

\begin{figure}[ht]
\centering
\includegraphics[width=0.48\textwidth]{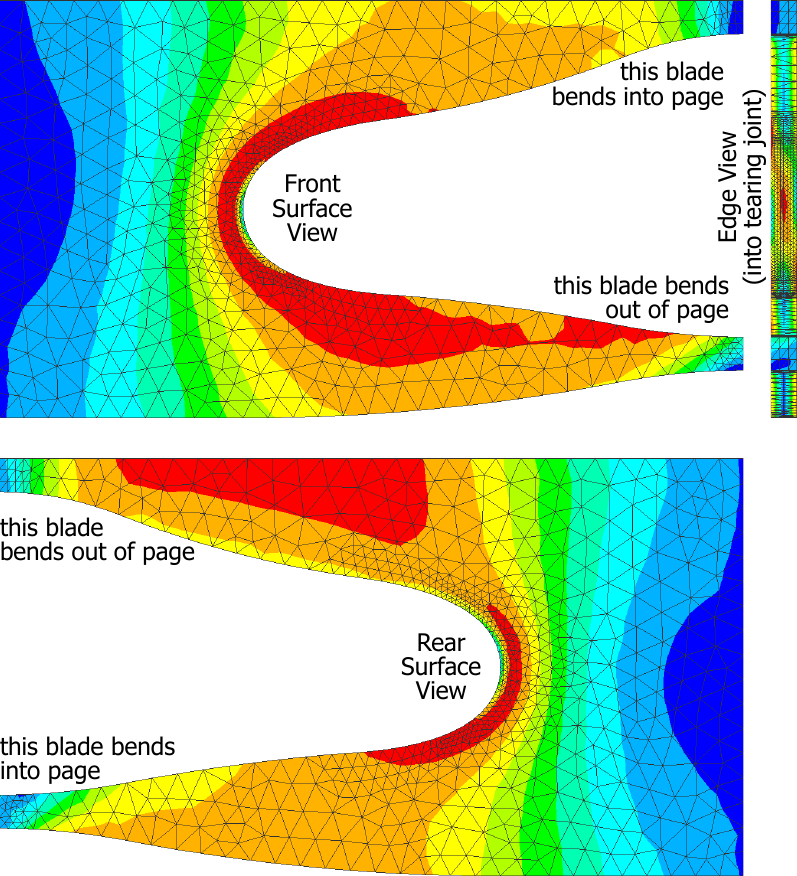}
\caption{Von-Mises stress level patterns on blade surfaces with optimal shaping showing that the tearing stresses are equal to the compressive bending stresses.}
\label{fig:StressPattern}
\end{figure}

Figure~\ref{fig:StressPattern} illustrates the most even stress distribution obtained with $n\leq1$ optimisation. As can be seen, the peak stress around the tearing joint is very similar to the peak stress along the concave (compressing) sides of the blades. This quarter blade length has been compressed by 0.8\,mm representing a compression of the entire spring by approximately 3.2\,mm. The peak strain in this analysis reaches 1.38\% which is beyond the capabilities of most spring materials but represents only 60\% of the maximum flexural capability of the material we used for the test in section~\ref{InstRes}.

With $n$ at higher values (making force more important than stress evenness), so that the blade width takes a larger fraction of the material width, then the force obtained increases as shown in Fig.~\ref{fig:OptValPlots}(a). This increase in force is not linear with increase in blade width (indicated by the dashed line) but increases with some higher power. The reason for this is not well understood but is expected to be related to the effective length of the blades shortening as the smaller gaps between blades provide tighter end-constraints than large gaps and because the Euler buckling force increases with decreasing length (going as length squared). As the (buckling) load force is almost independent of displacement, there is significant gain in operating in this regime when increased load force is more important than obtaining maximum displacement. We chose the shape obtained with $n=2$ for subsequent experimental measurements to obtain this increased load capability.

Another effect of the increased blade width fraction and force generation, is that the stress becomes concentrated around the tearing joints and takes on higher values than the simple-to-calculate blade bending stress that appears evenly distributed throughout for $n=1$. This effect is shown in Fig.~\ref{fig:OptValPlots}(b).  These peak stress values appearing around the tearing joints were obtained with the same 3.2\,mm compression and thus the same bending stress along the blade as was obtained for $n=1$. So to operate in this regime we need to de-rate the maximum compression distance so that the blade bending stress is kept this factor below its maximum yield value.  For example, for our choice of $n=2$, the spring blades at their widest place used 78\% of the material width, and so (reading from Fig.~\ref{fig:OptValPlots}(b) at 0.78) we needed to limit the bending stress in the blades to $1/1.23$ of its maximum yield value in order to ensure that the actual maximum was not exceeded around the tearing joints.

Fig.~\ref{fig:OptValPlots}(c)\&(d) indicate the optimum values obtained for the Bezier shape and Bezier extent dimensions. To normalize these values, the Bezier dimensions have been taken as a ratio to the radius of a simple fillet bridging the space between blades (see the yellow semicircle in Fig.~\ref{fig:AnsysGeom}(a)).  The scatter in these plots indicate that the values obtained are not a strong optimum when normalised in this manner.

\section{Dynamic response simulation}\label{Dynamic}

In order to obtain some indication of the dynamic effect of the reduction in blade mass obtained by shape contouring, the frequency response of a simple rectangular and very similarly rated contoured blade were simulated for comparison.  The contoured spring dimensions were as determined by optimising $F_{\rm{load}} / \sigma_{\rm{max}}$ in section~\ref{PracDes} and similar to the the sample that we describe static testing of in section~\ref{InstRes} (material $74.6\times25.4\times0.75$\,mm, blade width 61\%, blade length 60\,mm). The comparison rectangular spring had its blade width adjusted to achieve the same reaction force as the contoured blade and simple radius fillets were made between the blades to replace the contoured Bezier.

\begin{figure}[ht]
\centering
\includegraphics[width=0.48\textwidth]{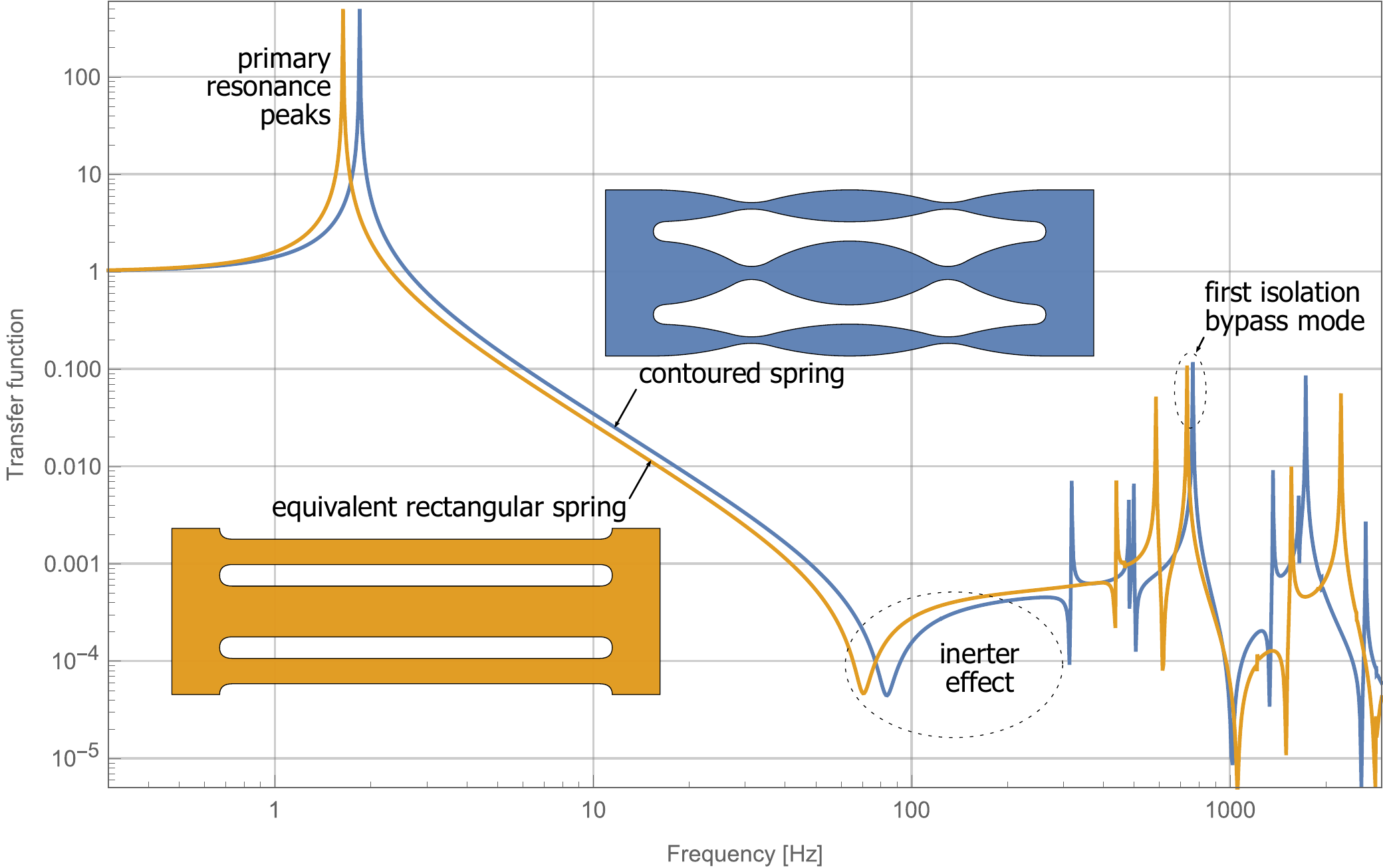}
\caption{Comparative simulation in ANSYS between a contoured and equivalent rectangular Euler spring.}\label{fig:FreqResp}
\end{figure}

The simulated transfer function of the two versions of the three-blade Euler springs in the 3.2\,mm post-buckled compressed state with a load of approximately 45\,kg are shown in Fig.~\ref{fig:FreqResp}. As expected, the primary resonance of the contoured spring was higher than the rectangular spring because of its higher spring-rate or shorter effective length (see table~\ref{table:EffLen}), and it provides $1/f^2$ isolation to 40 times its primary resonance. The notch and leveling after this initial slope is due to the fact that the spring's central mass must expand and accelerate out sideways at a much higher rate than the compression motion that causes it. As a result of this high gear ratio effect, the momentum of this moving mass acts as a small \emph{inerter} producing the isolation-limiting effect shown. As this is very similar to the often seen center of percussion effect, it can be cancelled by means of an appropriately mass loaded lever-arm with opposite centre of percussion as previously demonstrated in ref.~\cite{Winterflood_2002}.

The first internal mode of the contoured spring is significantly lower in frequency than its rectangular counterpart. But these modes are only parasitic pole-zero pairs which are unlikely to be prominent due to the loss coefficient in practical spring materials and could also be very easily damped with an inertial absorber. The first troublesome mode which would bypass vibration isolation at a transmissibility higher than 10\% appears between 700 and 800\,Hz for both spring types. This is a factor 400 higher than the resonance frequency, which can be compared to the factor 1000 for the pendulum horizontal isolator and 150 for the triangular cantilever blade presented earlier in Fig.~\ref{fig:SuspCompare}. Apart from the greatly increased working range of the contoured spring, there appears to be not much difference in isolation performance between the two shapes.

\section{Measurement of a glassy metal Euler spring}\label{InstRes}

The property that makes a material useful as a spring (if mass is not so critical) is its energy storage capability per unit volume. This figure of merit can be characterised by yield strength squared divided by modulus of elasticity. Comparing some high strength metal alloys using aluminium 7075T6 as reference gives beryllium-copper as 2.7 times better and maraging steel (2395\,MPa yield) as 8.1 times better. We also considered the glassy metal LM105 which has a merit factor 7.1 times better than AL7075 when the the bulk yield strength is used in the formula. However, since Euler springs are stressed in bending, and this material is significantly stronger under bending than under uniform tension, it is appropriate to use its flexural strength for this figure of merit.

\begin{figure}[ht]
\centering
\subfigure[~]{
\includegraphics[width=0.165\textwidth]{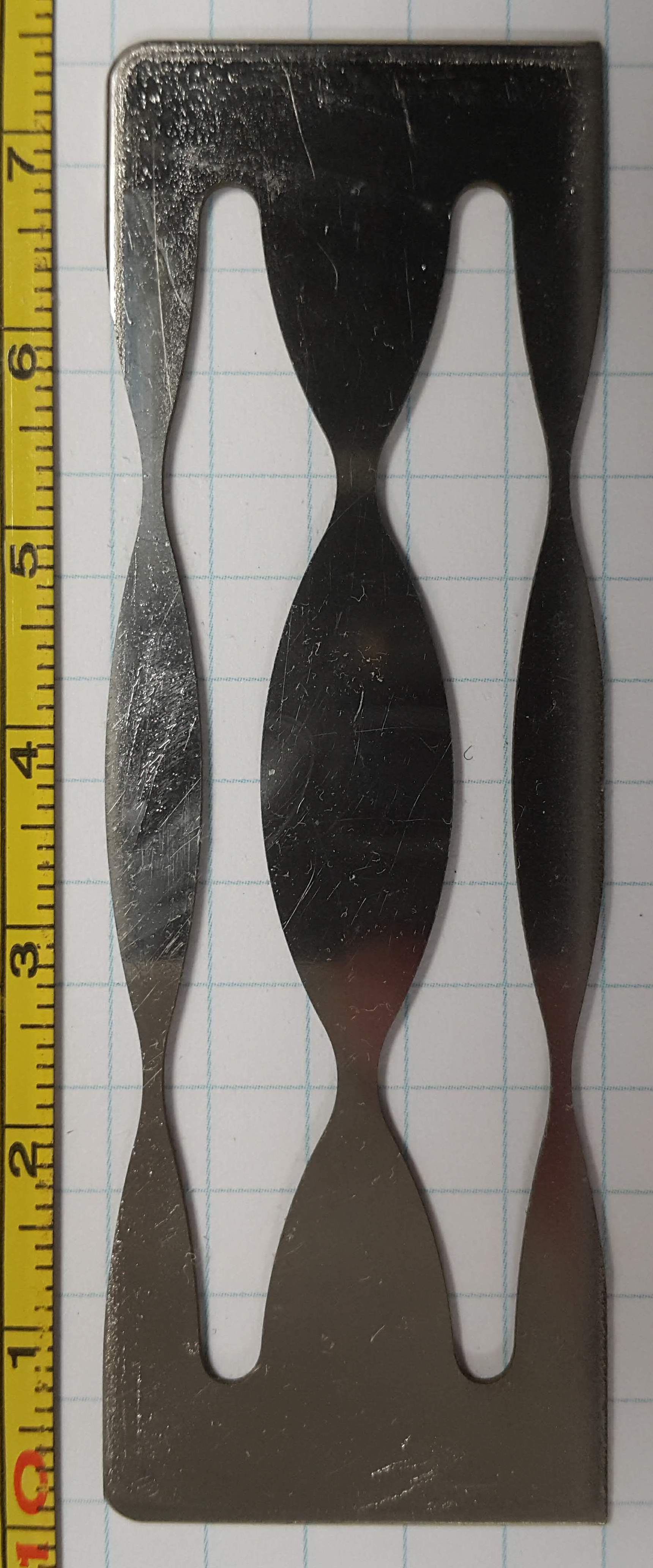}
\label{sfig:Photo3L}
}
\subfigure[~]{
\includegraphics[width=0.2715\textwidth]{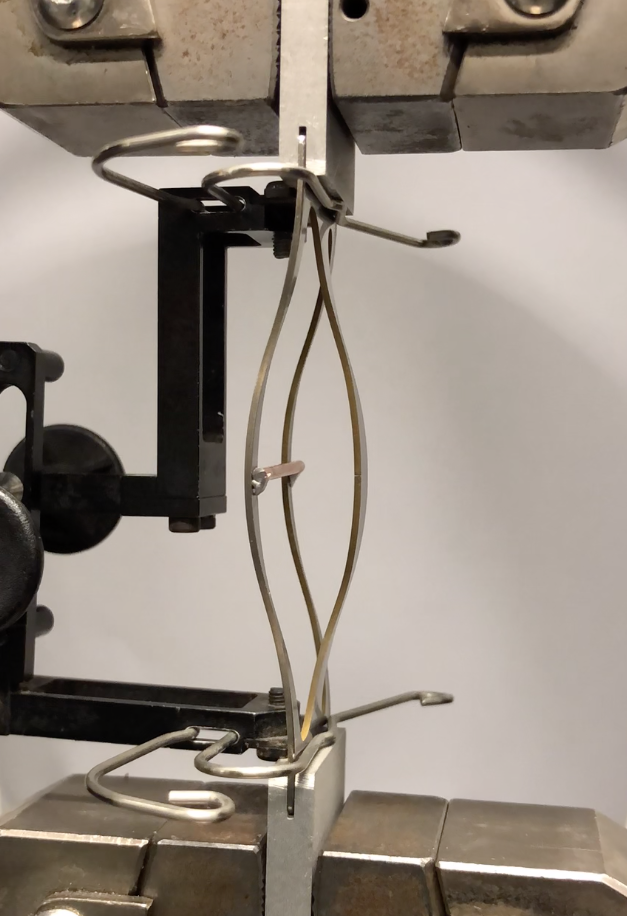}
\label{sfig:GMinCT}
}
\caption{Panel~\subref{sfig:Photo3L} shows a three blade profile optimised by ANSYS for $F_{\rm{load}}^2 / \sigma_{\rm{max}}$. Panel \subref{sfig:GMinCT} shows a still from analysis video of compression test of this glassy metal contoured Euler spring. Top and bottom show claws of tester with the compressed and buckled blade in between. A small cross bar is inserted between the three blades to assist in correct buckling direction. An extensometer to verify the Instron's ballscrew drive position is visible on the left. Full video found in ref.~\cite{Youtube_2019}. Cycles take about 20 seconds.}\label{fig:Photos}
\end{figure}

Using its flexural strength, the glassy metal is superior to all other spring materials that we considered with a merit factor 13.9 times better than AL7075. We expect the reason that the flexural strength is so much greater than its bulk yield strength is because the glassy state needs to be snap frozen from the molten state by very rapid cooling. This clearly happens fastest at the surface as it contacts the injection mould, and the surfaces are where all the bending stresses appear. Another advantage was that suitably sized samples were easy to obtain from liquidmetal.com, whereas sourcing a suitable quantity of maraging steel has proven very difficult.

A contoured monolithic three-blade Euler spring was cut from a sample sheet of LM105 glassy metal alloy 75x25x0.75\,mm, the dimensions of which were dictated by the size of samples available from Liquidmetal in March 2019. The shape of the blades and the tearing joints were determined from the optimisation discussed in section~\ref{PracDes} with $n=2$ and is shown in Fig.~\ref{fig:Photos}(a). Similar to previous tests, this spring was mounted in the Instron tester as shown in Fig.~\ref{fig:Photos}(b) and compression tested over a distance of 1.9\,mm in multiple successive cycles of compression and relaxation.

\begin{figure}[ht]
\centering
\includegraphics[width=0.48\textwidth]{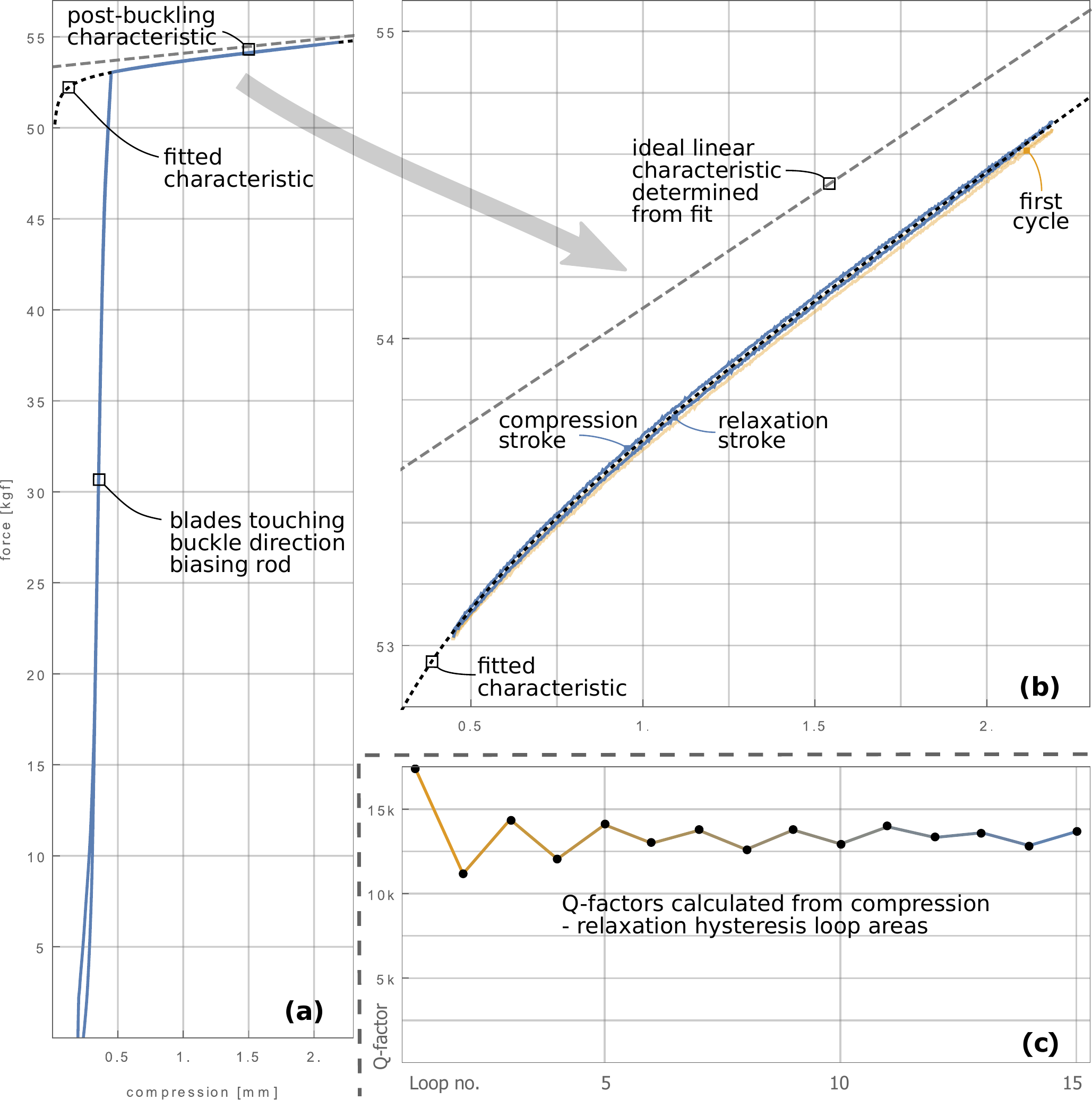}
\caption{Results from 8-cycle Instron compression testing. (a) Force-displacement characteristic of last cycle with origin included; (b) Expanded trace of free buckling segment with model fit; (c) $Q$-factors calculated from the area of the hysteresis loop between each adjacent compression or relaxation stroke.}
\label{fig:JorisResults}
\end{figure}

Results from one such compression test are shown in Fig.~\ref{fig:JorisResults}. Eight compression-relaxation cycles were logged and Fig.~\ref{fig:JorisResults}(a) shows the last cycle plotted on a scale which includes the origin. Fig.~\ref{fig:JorisResults}(b) shows first (buff) and last (blue) cycle of the set, expanded in scale, and only including those segments of the characteristic after the blades cease touching the direction biasing rod. As is evident from the first and last, all of the cycles looked very similar with no evidence of taking a significant set in the first cycle or of the set increasing during cycling. This may be because most setting would have occurred during setup and trials prior to data taking, and in this case we did not switch the blades to their opposite buckling direction for additional tests.

There is a clear 0.04\% increase in the critical load over the 8 cycles (from first to last) and we assume this was due to the air conditioning system bringing the room temperature down after recently being turned on and increasing the modulus of the material. This hypothesis is supported by the asymptotic approach of earlier cycles towards the last cycle. The extensometer reading was more non-linear than we would have liked (w.r.t the ballscrew crosshead position) so it was only used to determine the backlash ($\sim$1\,$\mu$m) in the crosshead positioning, and the backlash-corrected crosshead position used for all analysis and plotting.

As before, an estimate of the $Q$-factor can be made by the ratio of the energy loss per cycle to the stored energy as per equation~(\ref{eq:Qfact}) and these results are plotted in Fig.~\ref{fig:JorisResults}(c) for the 15 pairs of adjacent compression and relaxation strokes which close to form a hysteresis loop. The phase of the oscillation of the $Q$-factor between compression-relaxation vs relaxation-compression cycles is interesting because it is the opposite of what would be expected if more setting occurred during the compression stroke than the relaxation stroke. We believe this effect is due to the temperature settling to a lower value as the test progressed so that the second stroke of each pair showed a higher material modulus than the first.

As before, the characteristic curve can be accurately fitted using the three terms of Eq.~\ref{eq:Fit} (together with adjustable origin) and this fit is shown by the black dotted line. Keeping only linear terms by setting the eccentricity factor to zero gives the ideal linear characteristic shown in dashed gray. This fit indicates that, if the eccentricity could be removed, this spring would have a buckling load of 53.3\,kgf and a post-buckled spring coefficient of 0.746\,kgf/mm or 7.32\,kN/m.  This equates to an effective length (see Fig.~\ref{fig:ForceDisp}) of 71.4\,mm or $1.19\ell$ (using $\ell$=60\,mm slot length). This spring coefficient, combined with the mass loading required for post-buckling operation ($\sim$54\,kg), gives a potential resonance frequency of 1.85\,Hz. This effective length (and consequent resonance frequency) differs significantly from the $1.67\ell$ (1.57\,Hz for $\ell$=60\,mm) expected from the analysis of section~\ref{ShtoSt} and summarised in table~\ref{table:EffLen}. This large discrepancy is not well understood, but it is expected to be related to the shearing stresses in the tearing joints which are not considered in the simplistic analysis of section~\ref{ShtoSt} which assumes simple clamped-end bending across the roots of the tearing joints. This measured stiffness is more in agreement with the simulated frequency response shown in Fig.~\ref{fig:FreqResp} which shows a resonance frequency of 2\,Hz and this discrepancy also suggests that the analysis of section~\ref{ShtoSt} is too simplistic for accurate results.

\section*{Conclusions and future prospects}

It has become clear from our testing of side-by-side oppositely buckling blades that this technique easily achieves a text-book force-displacement buckling response with very little of the stick-slip degradation that plagued our previous clamped blade technique. The majority of the changes in force-displacement characteristic that appeared between the first and subsequent compression cycles in our early section~\ref{SbsTest} tests are likely due to not taking proper account of the high level of stress (likely exceeding yield) around the tearing joints. The wide slot requirements ($\sim$40\% of material width) indicated by the FEM analyses of section~\ref{PracDes} will avoid this deficiency in the future.

The optimal shape that contoured blades should be cut in order to most evenly distribute stress and minimize blade mass is here presented. It is clear from first principles that this shape must give the largest compression distance possible for a given blade length, material thickness and yield strength, with improvements typically in excess of 60\% over similar rectangular blades. This is because the maximum curvature is obtained over the full length of the blade (except for the narrow necks). The elastica by comparison has large sections of much reduced curvature.

If the optimal shape with $\sim$60\% of the material width used in the springs and $\sim$40\% used for the Bezier rounding between blades, does not provide sufficient force for some applications, then the compression distance and thus stress level can be de-rated from the blade bending maximum, according to the Relative stress curve in Fig.~\ref{fig:OptValPlots}.

Apart from the major advantage of significantly increased compression distance for a given stress level (or equivalently reduced stress level for a required compression range), contouring the blade shape has some slight disadvantages. The effective length is somewhat shorter giving a higher resonance frequency in typical applications. The spring coefficient is also more non-linear than its rectangular equivalent making it more difficult to reduce by combining it with some negative spring effect. There is very little difference between the internal modes of the two blade shapes for vibration isolation applications.

We expect that the next advance in performance of Euler buckling springs will be achieved by varying the thickness of the spring blades rather than the shape contour. This should allow monolithic type boundary conditions as well as allowing more of the spring mass to be concentrated in areas where spring motion is minimum. Monolithic in thickness boundaries have been made from plastic with 3D-printing\,\cite{virgin2021simultaneous} but blade thickness was kept uniform.

The materials tested produced some surprises. We were amazed at the high $Q$-factors obtained using razor blade steel and disappointed with the low values obtained with the glassy metal. We hope to further investigate this strangeness. Not reported are some trials we did using tungsten carbide for blade material. We hoped that its extra high compressive strength might be ideally suited to the dominantly compressive stresses appearing in Euler buckling. However, the large degrees of setting and hysteretic losses that were observed due to the softness of the cobalt binding matrix meant that it was totally unsuitable for this application.

\section*{Acknowledgements}
This work was funded by the Australian Research Council (ARC) Centre of Excellence for Gravitational Wave Discovery OzGrav under grant CE170100004. We would like to thank Andrew Sunderland, Alessandro Bertolini, David Blair, Bram Slagmolen and Joshua McCann for useful discussions and comments. Additionally, we would like to acknowledge contributions from students Lindsay Wood and Wenjing Zheng early on. The authors also wish to thank the workshop technicians Ken Field and Steve Key.

\bibliographystyle{unsrt}
\bibliography{refs}

\end{document}